\documentclass[twocolumn,aps,prl,superscriptaddress,amsmath,amssymb]{revtex4-1}
\usepackage{color}
\usepackage{amstext}
\usepackage{graphicx}
\usepackage{esint}

\makeatletter

\usepackage{xcolor}
\usepackage{subfigure}
\usepackage{dcolumn}
\usepackage{bm}
\usepackage{framed}
\usepackage{dsfont}
\usepackage[normalem]{ulem}
\usepackage{nicefrac}

\usepackage{nicefrac}
\usepackage{braket}

\newcommand{\be}{\begin{equation}}
\newcommand{\ee}{\end{equation}}
\newcommand{\bea}{\begin{eqnarray}}
\newcommand{\eea}{\end{eqnarray}}

\makeatother

\begin{document}
\title{Nonequilibrium fluctuations of  a quantum heat engine}
\author{Tobias Denzler}
\affiliation{Institute for Theoretical Physics I, University of Stuttgart, D-70550
Stuttgart, Germany}
\author{Jonas F.~G. Santos}
\affiliation{Centro de Ci\^{e}ncias Naturais e Humanas, Universidade Federal do ABC,
Avenida dos Estados 5001, 09210-580 Santo Andr\'{e}, S\~{a}o Paulo, Brazil}
\author{Eric Lutz}
\affiliation{Institute for Theoretical Physics I, University of Stuttgart, D-70550
Stuttgart, Germany}
\author{Roberto M. Serra}
\affiliation{Centro de Ci\^{e}ncias Naturais e Humanas, Universidade Federal do ABC,
Avenida dos Estados 5001, 09210-580 Santo Andr\'{e}, S\~{a}o Paulo, Brazil}
\begin{abstract}
The thermodynamic properties of quantum heat engines are stochastic  owing to the presence of thermal and quantum fluctuations. We here experimentally investigate the efficiency and nonequilibrium entropy production statistics of a spin-1/2 quantum Otto cycle. We first study the correlations between work and heat within a cycle by extracting their joint  distribution  for different driving times. We show that near perfect anticorrelation, corresponding to the tight-coupling condition, can be achieved. In this limit, the reconstructed   efficiency distribution is peaked at the macroscopic efficiency and fluctuations are strongly suppressed. We further test the second law  in the form of a joint fluctuation relation for work and heat. Our results characterize the statistical features of a small-scale thermal machine in the quantum domain and provide means to control them.
\end{abstract}
\maketitle

Heat engines have played a prominent role in  our society since the industrial revolution.
They are commonly used to generate motion by converting thermal energy into mechanical work \citep{cen01}. An important figure of
 merit of  heat engines is their efficiency, defined as the ratio
 of work output and heat input. According to the second law of thermodynamics, the maximum efficiency of any thermal  motor operating between two heat baths is given by the Carnot formula, $\eta_\text{ca} = 1- T_1/T_2$, where $T_{1,2}$ denote the respective temperatures of the cold and hot  reservoirs \citep{cen01}. For macroscopic heat engines consisting of a huge number of degrees of freedom, heat, work and, consequently, efficiency are deterministic quantities.

In the past decade, successful  miniaturization has led to the experimental downscaling of thermal machines to  microscopic \cite{sta11,bli12,mar15,pro16} and nanoscopic \cite{hug02,ros16,lin19} levels. Quantum heat engine operation has furthermore been reported recently in a variety of systems   \cite{zou17,kla19,ass19,pet19,hor20,bou20}. Contrary to macroscopic engines,  small motors  are subjected to  thermal \cite{sei12} and, at low enough temperatures, to  additional quantum \cite{esp09,cam11} fluctuations. These are associated with random transitions between discrete energy levels, and thus introduce nonclassical features. As a result, heat, work, efficiency, and other
relevant thermodynamic quantities such as the nonequilibrium entropy production,  are stochastic variables.  Such fluctuations strongly impact the performance of microscopic  and nanoscopic machines \cite{ver14,pol15,jia15,man19}. Understanding their random properties is therefore essential. The efficiency statistics of classical Brownian heat engines has been studied experimentally with optically trapped colloidal particles in Refs.~\cite{mar15,pro16}. Remarkably, efficiency fluctuations above the Carnot efficiency, which originate from negative entropy production events, have been observed \cite{mar15,pro16}. Meanwhile, the random entropy production for arbitrary heat engines has been theoretically predicted to satisfy a fluctuation relation \cite{sin11,lah12,cam14},  a fundamental nonequilibrium generalization of the second law  of thermodynamics for small systems \cite{sei12,esp09,cam11}.  However, the efficiency and nonequilibrium entropy production statistics  of  quantum heat engines have not been explored experimentally so far.

\begin{figure}[t]
\centering \includegraphics[width=0.48\textwidth]{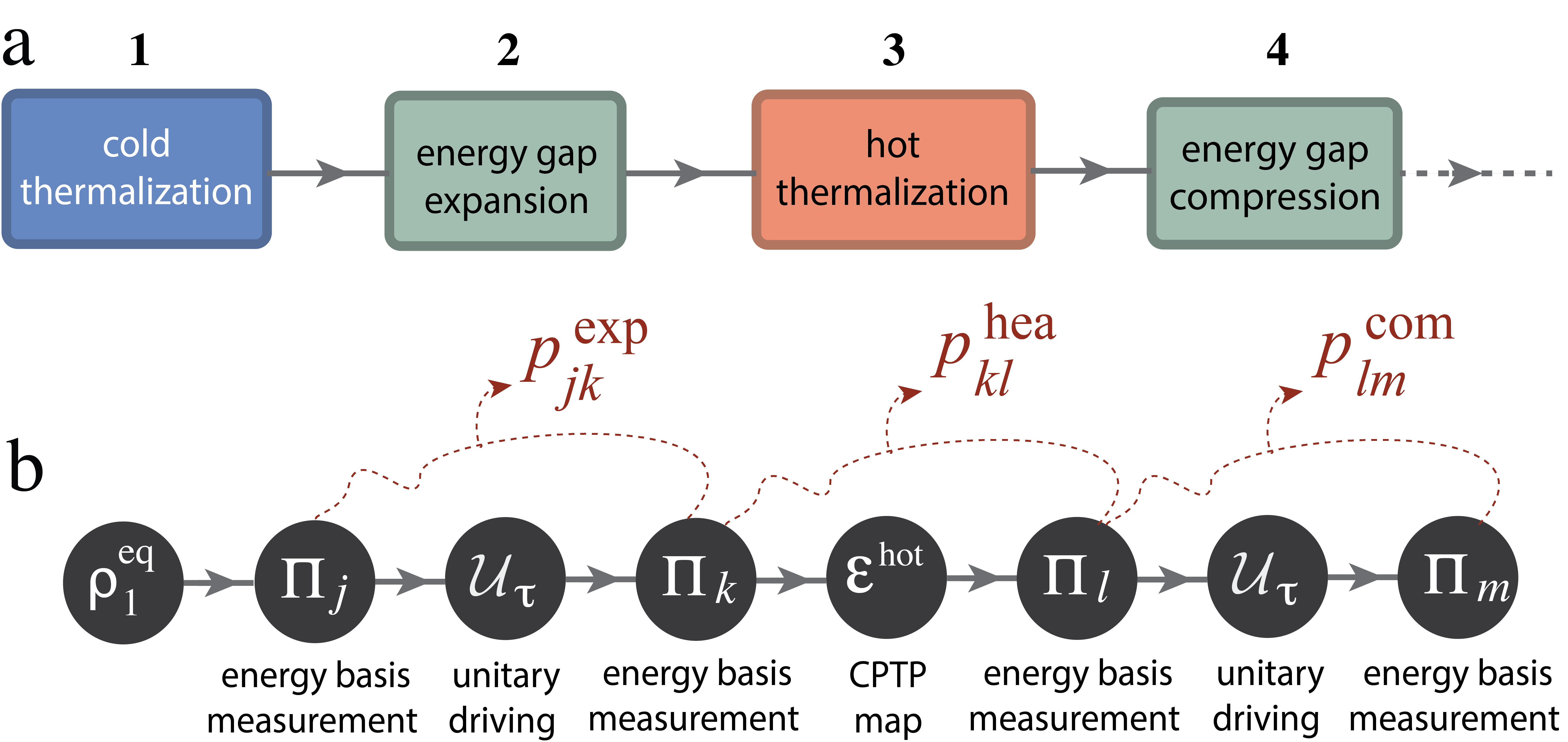} \caption{Quantum heat engine. (a) Four steps (cooling, unitary expansion, heating, unitary compression) of the quantum Otto cycle 
 realized  in the experiment. (b) Multi-point-measurement scheme used to determine the joint  distribution $P(W,Q)$ of work and heat:  projective energy measurements are performed at the beginning ($\Pi_j$) and at the end ($\Pi_k$) of the expansion stroke, as well as  at the beginning ($\Pi_l$) and at the end ($\Pi_m$) of the compression phase. Each pair of measurements is realized via a Ramsey-like interferometric method. The operator $\mathcal{U}_{\tau}$ describes unitary driving and  $\varepsilon^{\text{hot}}$ characterizes the completely positive trace preserving (CPTP) map that fully thermalizes the system to the hot temperature.
}
\label{fig:1}
\end{figure}

\begin{figure*}[ht!]
\centering \includegraphics[width=1\textwidth]{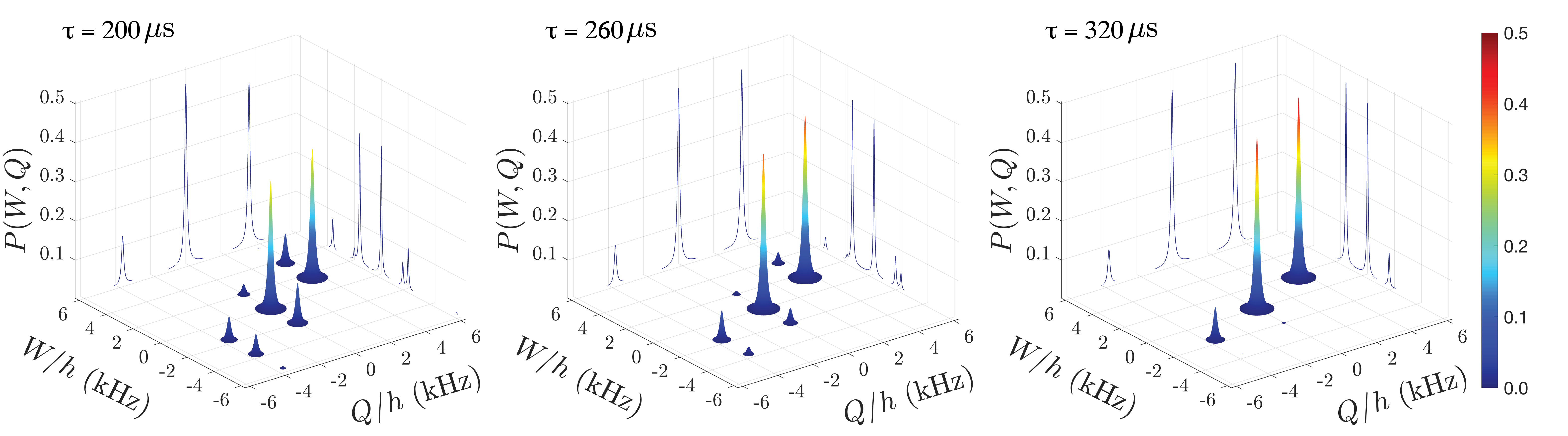} \caption{Joint  distribution of work and heat.  Distribution $P(W,Q)$, Eq.~(1), for three driving times, $\tau=200,\text{ }260,\text{ and }320$ $\mu$s
in the unitary driving strokes. Experimental data is well fitted by nine
Lorentz distributions at the corresponding pair of stochastic values
for the extracted work ($W/h$ = $0,\pm1.6,\pm2.0\pm3.6,\pm5.6$\textcolor{blue}{~}kHz)
and heat absorbed from the hot source ($Q/h$ = $0,\pm3.6$\textcolor{blue}{~}kHz). Diagonal peaks grow at the expense of off-diagonal ones as the quasiadiabatic regime is approached at $\tau=320\mu$s.}
\label{fig:2}
\end{figure*}

We here report the study of the fluctuating  properties of a quantum Otto engine \cite{kos17} based on a driven nuclear spin-1/2 in a liquid state nuclear magnetic resonance (NMR) setup \cite{oli07}. We extend existing interferometric methods \cite{dor13,maz13,bat14,bat15} to extract the joint  distribution of work and heat for different cycle times. We exploit the multipoint statistics to investigate  the correlations between work and heat during an engine cycle, from the adiabatic to the nonadiabatic regime. We find  near perfect anticorrelation, corresponding to the tight-coupling condition \cite{bro05,esp09a,cle15,sun19}, in the quasiadiabatic limit. We additionally determine the distribution of the quantum stochastic efficiency and analyze the impact of the work-heat correlations on its features. We show, in particular, that, as the tight-coupling regime is approached, the random  efficiency is peaked around the macroscopic  efficiency, and its fluctuations are strongly suppressed. We finally verify both a detailed and an integral bivariate quantum  fluctuation relation for cyclic heat engine operation \cite{sin11,lah12,cam14} that  has not been tested  before, and examine irreversible losses associated with  quantum friction \cite{kos02,kos03,zam14}.

In our experiment, we use  a $^{13}\textrm{C}$-labeled
$\textrm{CHCl}_{3}$ liquid sample diluted in Acetone-D6 and a $500$~MHz
Varian NMR spectrometer. We employ the spin 1/2 of the $^{13}\textrm{C}$ nucleus
as the working medium of the quantum  heat engine and the  $^{1}$H nuclei  as a heat bus to deliver
heat to the  machine. Work is performed by driving the heat engine with a resonant radio-frequency (rf) field. The low rf modes
near to carbon resonance frequency  further act as the cold heat bath, while
high rf modes near the hydrogen Larmor frequency operate as the hot reservoir.

We realize a  quantum Otto cycle that consists of four different steps \cite{kos17}
(Fig.~1a). 1) Cooling: the $^{13}$C
nuclear spin is initially cooled, using spatial average techniques \cite{oli07}, to a pseudo-thermal state $\rho_{1}^{\text{eq}}=\exp({-\beta_{1}H_{1}^{\text{C}}})/Z_{1}$ at  cold inverse spin temperature $\beta_{1}$, where $H_{1}^{\text{C}}$ is  the initial Hamiltonian and $Z_{1}$ the partition function. 2) Expansion: the machine is then driven by a  time-modulated rf field
on resonance with the $^{13}\textrm{C}$ nuclear spin. In a rotating
frame at the $^{13}\textrm{C}$ Larmor frequency ($\approx125~\text{MHz}$),
the driving is described by the following effective Hamiltonian, 
$
H_{\text{exp}}^{\text{C}}(t)=-(h\nu(t)/2)\left[\cos\left({\pi t}/{2\tau}\right)\sigma_{x}^{\text{C}}+\sin\left({\pi t}/{2\tau}\right)\sigma_{y}^{\text{C}}\right],
$
where the nuclear spin energy gap, $h\nu(t)=h\nu_{1}\left(1-{t}/{\tau}\right)+h\nu_{2}{t}/{\tau}$
is varied linearly from $\nu_{1}=2.0$\textcolor{blue}{~}kHz at
time $t=0$ to $\nu_{2}=3.6$\textcolor{blue}{~}kHz at time $t=\tau$,
where $\sigma_{x,y,z}^{\text{C}}$ are the Pauli spin operators
of the $^{13}\textrm{C}$ nuclear spin. 
Implemented driving times ($\approx10^{-4}$~s) are much shorter than the typical decoherence
times in our setup (few seconds), implying that the corresponding evolution  $\mathcal{U}_{\tau}$ is unitary to an excellent approximation \cite{bat14}.
3) Heating: heat exchange between the $^{13}\textrm{C}$ and the $^{1}\textrm{H}$ nuclear spins, which was prepared at
 the hot inverse temperature $\beta_2$ \cite{mic19},  is achieved by a sequence of free evolutions under the natural
scalar interaction $H_{J}=({\pi}/{2})hJ\sigma_{z}^{\text{H}}\sigma_{z}^{\text{C}}$
(with $J\approx215.1~\text{Hz}$) between both nuclei and rf pulses \cite{pet19} (Supplementary Information). The resulting fully thermalized state is $\rho_{2}^{\text{eq}}=\exp({-\beta_{2}H_{2}^{C}})/Z_{2}$, with $H_{2}^{C}=H_{\text{exp}}^{\text{C}}(\tau)$. 4) Compression: we finally decrease the nuclear spin energy gap back to its initial value $\nu_1$ in time $\tau$ according to $H_{\text{com}}^{\text{C}}(t)=- H_{\text{exp}}^{\text{C}}(\tau-t)$. In both cases, $\beta_{i}=(k_{B}T_{i})^{-1}$, ($i= 1, 2$), where $T_i$ is the spin temperature and $k_B$ the Boltzmann constant.

\begin{figure*}[t]
\centering \includegraphics[width=1\textwidth]{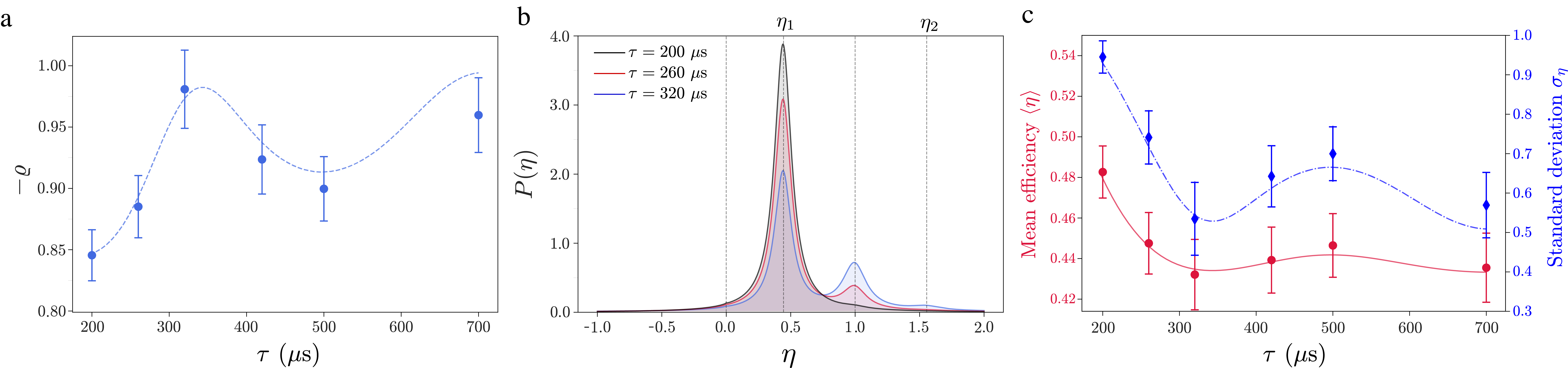} \caption{Work-heat correlations and efficiency distribution. (a) Pearson
correlation coefficient $\varrho$ for work and heat as a function of the
 driving time $\tau$.  (b)
Efficiency distribution $P(\eta)$, Eq.~(2), for  three
different driving times, displaying two large peaks at  1 and $\eta_1 = \eta_\text{th}\approx 0.44$, and two small peaks at  0 and $\eta_2= 2-\eta_\text{th}\approx1.56$.
(c) Mean microscopic efficiency $\langle \eta \rangle$ and corresponding  standard deviation $\sigma_\eta$  as a function of the driving time $\tau$. Work and heat are maximally anticorrelated as the tight-coupling condition is approached for $\tau = 320\mu$s.  As a result, the stochastic efficiency is sharply peaked around the macroscopic efficiency $\eta_\text{th}$ and  the standard deviation, which quantifies  fluctuations, strongly decreases. By contrast, the mean microscopic efficiency decreases. Symbols are the experimental data
and   lines show theoretical predictions.}
\label{fig:3}
\end{figure*}

Work and heat fluctuations of the quantum heat  engine are characterized by a joint distribution $P(W,Q)$, which can be fully determined in the present
experiment by a multipoint measurement scheme along the quantum Otto
cycle for different driving times $\tau$ (Fig.~1b). The protocol consists of two projective energy measurements at the beginning ($\Pi_j$) and at the end ($\Pi_k$) of the expansion stroke, as well as two additional projective energy measurements at the beginning ($\Pi_l$) and at the end ($\Pi_m$) of the compression phase. Each of the three consecutive pairs of measurements is realized via a Ramsey-like interferometric method \cite{dor13,maz13,bat14,bat15} and allows the determination of the respective transition probability. The corresponding joint   distribution of the total extracted work $W$ and the absorbed heat $Q$ reads (Supplementary Information),
\begin{eqnarray}
P(W,Q) & = & \sum_{j,k,l,m}\Delta(W,j,k,l,m,\tau,\gamma)\Delta(Q,j,k,l,m,\tau,\gamma)\nonumber \\
 & &\times  p_{j}^{0}\,p_{jk}^{\text{exp}}\,p_{kl}^{\text{hea}}\,p_{lm}^{\text{com}},
\end{eqnarray}
where $p_{j}^{0}={\exp({-\beta_{1}E_{j}^{0}})}/{Z^{0}}$
is the  occupation of the cold equilibrium state, with $E_{j}^{0}$
eigenenergies of $H_{1}^{\text{C}}$. The transition probabilities during expansion, heating and compression are respectively  $ p_{jk}^{\text{exp}}$, $p_{kl}^{\text{hea}}$ and $p_{ml}^{\text{com}}$. Since the heating stroke leads to a hot equilibrium state, we simply have $p_{kl}^{\text{hea}}=p_{l}^{\tau}={\exp({-\beta_{2}E_{l}^{\tau}})/}{Z^{\tau}}$, independent of $k$, with $E_{j}^{\tau}$  eigenenergies of $H_{2}^{\text{C}}$. Occupation probabilities describe the effects of thermal fluctuations, while
transition probabilities  those of quantum
fluctuations and quantum dynamics \cite{jar15}. For ideal projective measurements, each spectral peak is infinitely sharp ($\gamma=0$), and energy changes during single strokes are given by differences of energy eigenvalues \cite{tal07}. In this case, the two functions $\Delta$ associated with work and heat, $X=(W, Q)$,
are $\Delta(X,j,k,l,m,\tau,0)=\delta\left(X-x_{jklm} \right)$, with $w_{jklm} =E_{j}^{0} - E_{k}^{\tau}+E_{m}^{\tau} -E_{l}^{0}$  and $q_{jklm}=E_{l}^{\tau}-E_{m}^{\tau}$.
 However,  the experimental Ramsey-like interferometric scheme  leads to spectral peaks with a finite width $\gamma$, which are well fitted by a Lorentzian distribution, $\Delta(X,j,k,l,m,\tau,\gamma)=1/\{\pi\gamma[1+ (X-x_{jklm})^2/\gamma^2]\}$ \cite{bat14,bat15}.

Examples of experimentally reconstructed bivariate  distributions for work and heat  are shown in Fig.~2 for three different driving times, for $k_{B}T_{1}/h=1.60\pm0.02$\textcolor{blue}{~}kHz and $k_{B}T_{2}/h=12.21\pm0.89$\textcolor{blue}{~}kHz (results for additional driving times are presented in the Supplementary Information). We observe up to nine discrete Lorentzian peaks, each with a width of about $0.15$~kHz. As the driving time increases from $\tau = 200\mu$s to $\tau = 320\mu$s, diagonal peaks grow at the expense of off-diagonal ones. This suggests that work-heat correlations are enhanced as  the process becomes more and more adiabatic. 

Work-heat correlations within the heat engine cycle are conveniently studied quantitatively with the help of the Pearson coefficient, $\varrho= \text{cov}(W,Q)/\sigma_W\sigma_Q$, defined as the ratio of the covariance and the respective standard deviations \cite{bar89}. Work and heat are in general (strongly) anticorrelated ($\varrho<0$) for the quantum Otto engine (Fig.~3a) and correlations oscillate as a function of time, owing  to the periodic nature of the  driving during expansion and compression steps; dots represent experimental data and the dashed line a numerical simulation  (Supplementary Information). Maximum anticorrelation ($\varrho\simeq -1$) is achieved for quasiadiabatic driving for $\tau = 320\mu$s. In this limit, the quantum heat engine satisfies the  tight-coupling condition \cite{bro05,esp09a,cle15,sun19}, which implies that work and heat  are proportional to each other. The tight-coupling condition plays a special role in the investigation of the universal properties of heat engines \cite{bro05,esp09a,cle15,sun19}.

\begin{figure}[t]
\centering \includegraphics[width=0.48\textwidth]{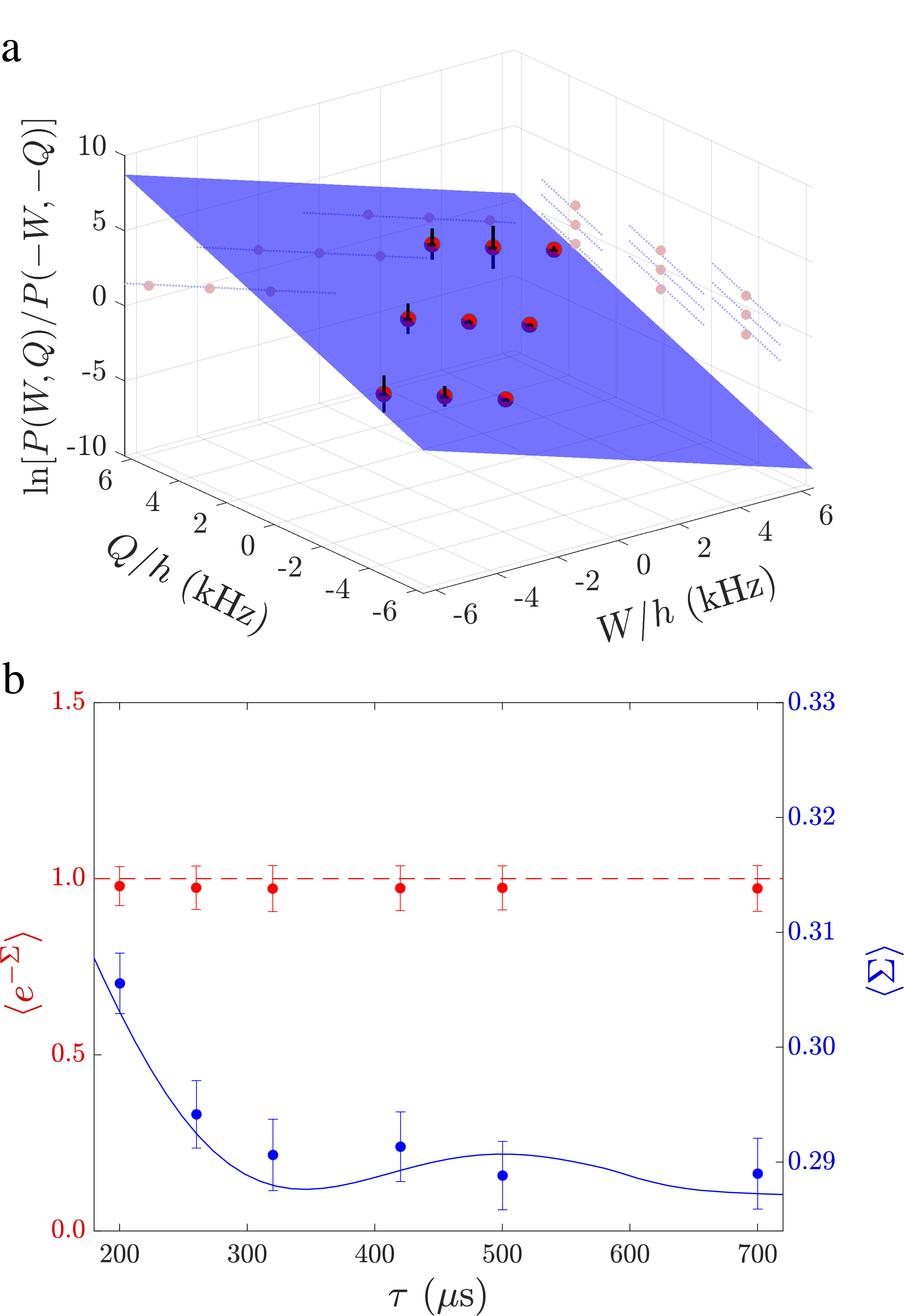} \caption{Nonequilibrium quantum fluctuation relations. (a) Verification
of the detailed fluctuation relation (3) for the quantum Otto cycle:  the values of $\ln\left[P(W,Q)/P(-W,-Q)\right]$ (red dots)
should lie within  the (blue) plane defined by the total entropy production $z= \Sigma =\Delta\beta Q-\beta_{1}W$; the dashed blue lines show the respective projections of the plane on the  work and heat axes. (b) Confirmation of the integral fluctuation theorem, $\langle \exp(-\Sigma)\rangle =1$, (red dots) and average entropy production $\langle \Sigma\rangle$ (blue dots) as a function of the driving time $\tau$. Irreversible losses are minimal when the tight-coupling condition is approached for $\tau = 320\mu$s.}
\label{fig:4-1}
\end{figure}

We next move to the analysis of the quantum stochastic efficiency defined as $\eta= W/Q$ for each single realization  \cite{den20}. This (random) quantity should not be confused with the (deterministic) thermodynamic efficiency,   $\eta_\text{th}=\langle W\rangle/\langle Q\rangle$, which is given in terms of averages  \cite{cen01}. In the case of adiabatic driving, the latter reduces to the standard Otto efficiency, 
$\eta_\text{Otto}=1-\nu_{1}/\nu_{2}$ \cite{kos17}. The efficiency distribution $P(\eta)$ follows from the joint distribution (1) 
via integration over all  work and heat values, 
\begin{equation}
P(\eta)=\iint dWdQ~\delta\left(\eta-\frac{W}{Q}\right)P(W,Q).\label{eq: P_eta}
\end{equation}
The corresponding experimental distribution is displayed in Fig.~3b for three different driving times. We identify four Lorentzian-like peaks:  two (large) peaks found at  1 and $\eta_1 = \eta_\text{th}$, and  two (small) peaks located at 0  and $\eta_2 = 2-\eta_\text{th}$(Supplementary Information). The stochastic efficiency is further seen to take values above 1 and below 0. In the former case, the produced random work is larger than the absorbed stochastic heat, while in the latter case work is  added to the machine or heat is given to the hot bath. These results indicate that all values of the stochastic efficiency are possible in a small-scale quantum engine running in finite time, including those forbidden by the macroscopic second law. As the driving time approaches the adiabatic regime ($\tau = 320\mu$s), we observe that  the macroscopic efficiency $\eta_\text{th}$ becomes increasingly more likely. In order to  examine the properties of the random microscopic efficiency $\eta$, we evaluate its mean $\langle \eta\rangle$ and standard deviation $\sigma_\eta$ in the interval $[-5,5]$ (Fig.~3c). The behavior of the mean efficiency and of its standard deviation as a function of $\tau$ is exactly opposite to that  of the Pearson coefficient (Fig.~3a): they decrease when the correlations increase, and vice versa, revealing the strong relationship existing between the work-heat (anti)correlations and the features of the stochastic efficiency. Surprisingly,  the dependence of the mean microscopic efficiency  $\langle \eta\rangle$ on $\tau$   is at variance with that of the macroscopic efficiency $\eta_\text{th}$ (Supplementary Information). The microscopic efficiency is thus larger for nonadiabatic driving than for adiabatic driving; this is due to the peaks above $\eta_\text{th}$ and, thus, to events violating the macroscopic second law which are more likely for nonadiabatic driving. We also  note that  the stochastic efficiency tends to the (deterministic) macroscopic  efficiency $\eta_\text{th}$ as the tight-coupling limit is approached. The effects of  fluctuations are here significantly suppressed due to the strong work-heat anticorrelation, even though these fluctuations do not vanish \cite{den20}.

Energy fluctuations in a heat engine cycle are predicted to obey a detailed fluctuation relation of the form  \citep{sin11,lah12,cam14},
\begin{equation}
\frac{P(W,Q)}{P(-W,-Q)}=e^{\Delta\beta Q-\beta_{1}W},\label{eq:Fluc_rel}
\end{equation}
where $\Delta\beta=\beta_{1}-\beta_{2}$ and $P(-W,-Q)$ is the joint 
distribution of measuring $(-W,-Q)$ in the reverse operation of the   engine. An integral  fluctuation theorem, $\langle \exp(-\Sigma)\rangle= \iint dWdQ~P(W,Q)\exp({-\Sigma})=1$, for the entropy production $\Sigma = \Delta\beta Q-\beta_{1}W$ follows after integration over one cycle \cite{sin11,lah12,cam14}.  The latter expression may be regarded as a nonequilibrium generalization of the Carnot formula, $\langle W\rangle /\langle Q\rangle \leq 1-T_1/T_2$, which can be derived from it by applying Jensen's inequality \cite{sin11,lah12,cam14}.
Figure \ref{fig:4-1}a displays
a verification of the  quantum detailed fluctuation relation (\ref{eq:Fluc_rel}) for $\tau=200$ $\mu$s
(see Supplementary Information for other  driving times).
We witness very  good agreement between the experimental values of $\ln\left[P(W,Q)/P(-W,-Q)\right]$ (red dots)  and
the predictions of Eq.~(\ref{eq:Fluc_rel}) indicated by the (blue) plane $z=\Sigma$, the $z$-axis being  vertical. A confirmation of the integral fluctuation theorem, $\langle \exp(-\Sigma)\rangle=1$, is further shown in Fig.~4b, as a function of the driving time,  together with the average entropy production $\langle \Sigma \rangle$, which characterizes irreversible losses within the  cycle.
Since the two driving Hamiltonians, $H_{\text{exp}}^{\text{C}}(t)$ and $H_{\text{com}}^{\text{C}}(t)$, do not commute at different times, the heat engine  exhibits internal friction associated with nonadiabatic transitions between
the instantaneous eigenstates of the $^{13}\textrm{C}$ nuclear spin \cite{kos02,kos03,zam14}. This purely quantum friction mechanism is the source of irreversibility in the quantum Otto cycle,  depending on the driving speed:  the mean entropy production  $\langle \Sigma\rangle$ decreases  as the adiabatic regime is approached (Fig.~4b), and vice versa. We also note a marked connection between  quantum friction (Fig.~4b) and work-heat correlations (Fig.~3a), which has not been acknowledged before.

In conclusion, we have performed the first experimental study of the work-heat correlations and their strong impact on both  efficiency and  entropy production statistics of a quantum heat engine. We have shown that the tight-coupling condition, corresponding to maximum work-heat anticorrelation, can be reached for finite-time quasiadiabatic driving. In this regime, the stochastic efficiency reduces to the macroscopic  efficiency, and both thermal and quantum fluctuations are notably suppressed. We have additionally observed that  macroscopic and microscopic efficiencies display opposite behavior, due to random events violating the macroscopic second law. We have finally confirmed  nonequilibrium generalizations of the Carnot formula in the form of bivariate fluctuation relations for work and heat, and analyzed the effect of quantum friction on the total entropy production. Our findings give a unique insight into the nonequilibrium fluctuating properties of small quantum thermal machines and provide direct means to control them.

\global\long\def\beginsupplement{%
\setcounter{table}{0} \global\long\global\long\global\long\def\thetable{S\Roman{table}}
\setcounter{figure}{0} \global\long\global\long\global\long\def\thefigure{S\arabic{figure}}
\setcounter{equation}{0} \global\long\global\long\global\long\def\theequation{S\arabic{equation}}
}
\clearpage
\section{Supplemental Materials}

This Supplementary Information provides additional details on the experimental protocol, the determination of the joint distribution for  work and heat, and the analysis of the experimental data. 

\subsection{Thermal states initialization}

Spatial average techniques \cite{oli07,bat14,bat15,mic19} were used to initialize the engine states,
which are local pseudo-thermal states encoded in the $^{1}$H and
$^{13}$C nuclei. We present in  Table \ref{tab:1}  the populations
and the respective local spin temperatures in the eigenbasis of Hamiltonians $\mathcal{H}_{0}^{\text{H}}$
and $\mathcal{H}_{0}^{\text{C}}$. 

\begin{table}[h]
\begin{centering}
\begin{tabular}{c|ccc}
$^{1}$H nucleus & $p_{0}^{\text{H}}$ & $p_{1}^{\text{H}}$ & $k_{B}T_{2}$ (peV)\tabularnewline
\hline 
 & 0.67 $\pm$ 0.01 & 0.33 $\pm$ 0.01 & 21.5 $\pm$ 0.4\tabularnewline
$^{13}$C nucleus & $p_{0}^{\text{C}}$ & $p_{1}^{\text{C}}$ & $k_{B}T_{1}$ (peV)\tabularnewline
\hline 
 & 0.78 $\pm$ 0.01 & 0.22 $\pm$ 0.01 & 6.6 $\pm$ 0.1\tabularnewline
\end{tabular}
\par\end{centering}
\caption{Populations and spin temperatures of the initial states of the  $^{1}$H and $^{13}$C nuclei.  The corresponding off-diagonal  elements  are zero within the measurement errors.}
\label{tab:1}
\end{table}

\subsection{Compression and expansion  protocols}

The energy gap compression and expansion protocols are implemented with a time-modulated amplitude and  phase transverse rf-pulse on resonance with the $^{13}$C nuclear spin in order to produce effectively  the time-dependent driving Hamiltonian $\mathcal{H}^{\text{C}}({\tau})$ described in the main text. The intensities of the transverse field at the beginning and end of the driving protocol were properly calibrated in order to have the associated frequencies given in the main text. The duration of the modulated traverse pulse was varied from $100$~$\mu$s to $700$~$\mu$s in different implementations of the quantum heat engine cycle.

\subsection{Heating protocol}

\begin{figure}[t]
\centering \includegraphics[width=0.98\columnwidth]{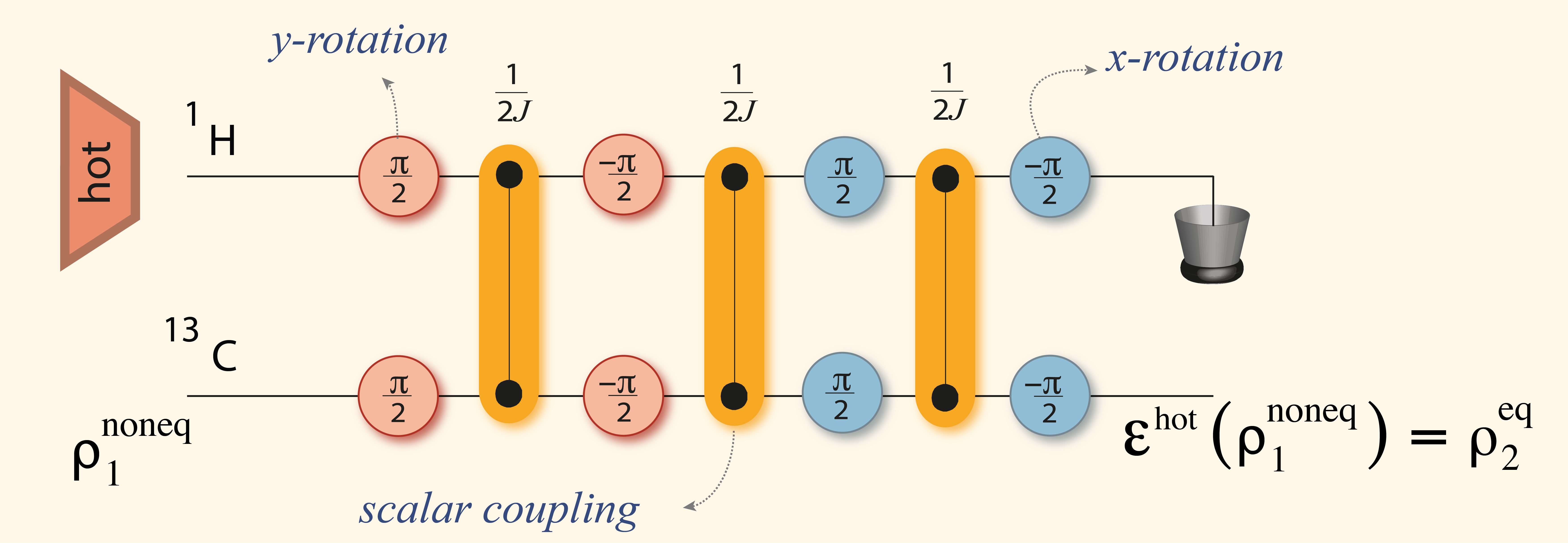} \caption{NMR pulse
sequence used in the heat exchange protocol. The outcome of this sequence
(which takes about $7${~}ms) is an effective full
thermalization described by a completely positive trace preserving
(CPTP) map on  reduced density operator of the carbon nucleus, $\varepsilon^{\text{hot}}:\rho_{1}^{\text{noneq}}\rightarrow e^{-\beta_{2}H_{2}^{\text{C}}}/Z_{2}$,
leading it to an equilibrium state at the hot inverse temperature
$\beta_{2}$. Orange connections represent free evolutions under the
scalar interaction during the time displayed above the symbol. Blue
(red) circles stand for $x$ ($y$) rotations by the displayed angle
implemented by transverse rf pulses.}
\label{figs1a}
\end{figure}

The  thermalization process used to heat the $^{13}$C nuclear spin of the quantum Otto engine during the second stroke has
  the local effect of a linear non-unitary
map $\varepsilon(\rho_{j})=\text{Tr}_{k\neq j}\left[\mathcal{U}_{\tau}\left(\rho_{H}^{0}\otimes\rho_{C}^{0}\right)\mathcal{U}_{\tau}^{\dagger}\right]$, with $(j,k) = (H,C)$.
 It is represented by the following
set of maps \cite{mic19}
\begin{equation}
\varepsilon(\rho_{j})=\sum_{\ell=1}^{4}K_{\ell}\rho_{j}^{0}K_{\ell}^{\dagger},
\end{equation}
 with the Kraus operators
\begin{align}
K_{1} & =\sqrt{1-p}\left(\begin{array}{cc}
1 & 0\\
0 & 0
\end{array}\right),K_{2}=\sqrt{p}\left(\begin{array}{cc}
0 & 0\\
0 & 1
\end{array}\right)\\
K_{3} & =\sqrt{1-p}\left(\begin{array}{cc}
0 & 1\\
0 & 0
\end{array}\right),K_{4}=\sqrt{p}\left(\begin{array}{cc}
0 & 0\\
-1 & 0
\end{array}\right).
\end{align}
 The parameter $p$ denotes the population of the excited state in the Hydrogen nucleus. The above Kraus operators correspond to the generalized amplitude damping of a single-$1/2$ system. From a local point of view, the map thus implements complete thermalization. The NMR pulse sequence used in the heat exchange protocol in the experiment is shown in Fig.~\ref{figs1a}, were the Hydrogen  nucleus is used as a heat bus.

\begin{table}[t]
\begin{tabular}{c|c|c|c|c|c|c}
History & stroke & stroke & stroke & stroke & $W/h\,\pm\,0.15$ & $Q/h\,\pm\,0.15$\tabularnewline
 & 1 & 2 & 3 & 4 & (kHz) & (kHz)\tabularnewline
\hline 
1 & $|\Psi_{-}^{1}\rangle$ & $|\Psi_{-}^{2}\rangle$ & $|\Psi_{-}^{2}\rangle$ & $|\Psi_{-}^{1}\rangle$ & 0 & 0\tabularnewline
\hline 
2 & $|\Psi_{-}^{1}\rangle$ & $|\Psi_{-}^{2}\rangle$ & $|\Psi_{-}^{2}\rangle$ & $|\Psi_{+}^{1}\rangle$ & -2.0 & 0\tabularnewline
\hline 
3 & $|\Psi_{-}^{1}\rangle$ & $|\Psi_{-}^{2}\rangle$ & $|\Psi_{+}^{2}\rangle$ & $|\Psi_{-}^{1}\rangle$ & 3.6 & 3.6\tabularnewline
\hline 
4 & $|\Psi_{-}^{1}\rangle$ & $|\Psi_{-}^{2}\rangle$ & $|\Psi_{+}^{2}\rangle$ & $|\Psi_{+}^{1}\rangle$ & 1.6 & 3.6\tabularnewline
\hline 
5 & $|\Psi_{-}^{1}\rangle$ & $|\Psi_{+}^{2}\rangle$ & $|\Psi_{-}^{2}\rangle$ & $|\Psi_{-}^{1}\rangle$ & -3.6 & -3.6\tabularnewline
\hline 
6 & $|\Psi_{-}^{1}\rangle$ & $|\Psi_{+}^{2}\rangle$ & $|\Psi_{-}^{2}\rangle$ & $|\Psi_{+}^{1}\rangle$ & -5.6 & -3.6\tabularnewline
\hline 
7 & $|\Psi_{-}^{1}\rangle$ & $|\Psi_{+}^{2}\rangle$ & $|\Psi_{+}^{2}\rangle$ & $|\Psi_{-}^{1}\rangle$ & 0 & 0\tabularnewline
\hline 
8 & $|\Psi_{-}^{1}\rangle$ & $|\Psi_{+}^{2}\rangle$ & $|\Psi_{+}^{2}\rangle$ & $|\Psi_{+}^{1}\rangle$ & -2.0 & 0\tabularnewline
\hline 
9 & $|\Psi_{+}^{1}\rangle$ & $|\Psi_{-}^{2}\rangle$ & $|\Psi_{-}^{2}\rangle$ & $|\Psi_{-}^{1}\rangle$ & 2.0 & 0\tabularnewline
\hline 
10 & $|\Psi_{+}^{1}\rangle$ & $|\Psi_{-}^{2}\rangle$ & $|\Psi_{-}^{2}\rangle$ & $|\Psi_{+}^{1}\rangle$ & 0 & 0\tabularnewline
\hline 
11 & $|\Psi_{+}^{1}\rangle$ & $|\Psi_{-}^{2}\rangle$ & $|\Psi_{+}^{2}\rangle$ & $|\Psi_{-}^{1}\rangle$ & 5.6 & 3.6\tabularnewline
\hline 
12 & $|\Psi_{+}^{1}\rangle$ & $|\Psi_{-}^{2}\rangle$ & $|\Psi_{+}^{2}\rangle$ & $|\Psi_{+}^{1}\rangle$ & 3.6 & 3.6\tabularnewline
\hline 
13 & $|\Psi_{+}^{1}\rangle$ & $|\Psi_{+}^{2}\rangle$ & $|\Psi_{-}^{2}\rangle$ & $|\Psi_{-}^{1}\rangle$ & -1.6 & -3.6\tabularnewline
\hline 
14 & $|\Psi_{+}^{1}\rangle$ & $|\Psi_{+}^{2}\rangle$ & $|\Psi_{-}^{2}\rangle$ & $|\Psi_{+}^{1}\rangle$ & -3.6 & -3.6\tabularnewline
\hline 
15 & $|\Psi_{+}^{1}\rangle$ & $|\Psi_{+}^{2}\rangle$ & $|\Psi_{+}^{2}\rangle$ & $|\Psi_{-}^{1}\rangle$ & 2.0 & 0\tabularnewline
\hline 
16 & $|\Psi_{+}^{1}\rangle$ & $|\Psi_{+}^{2}\rangle$ & $|\Psi_{+}^{2}\rangle$ & $|\Psi_{+}^{1}\rangle$ & 0 & 0\tabularnewline
\end{tabular}
\caption{All transition histories  between the instantaneous
eigenstates $|\Psi_\pm^i\rangle$ $(i=1,2)$ for each stroke of the   heat engine, together with the corresponding values of work and heat.}
\label{table2}
\end{table}

\subsection{Joint distribution for work and heat - theory}
The joint distribution for the total work $W$ and the absorbed heat $Q$ may be determined by performing energy  measurements on the engine at the beginning and at the end of the expansion, heating and compression strokes \cite{den20}, as depcted in Fig. 1\textbf{b} of the main text.
We first consider the case of ideal projective measurements. By performing projective energy measurements  at the beginning and at the end of the expansion step, the distribution of the expansion work $W_2$ reads \cite{tal07},
\begin{equation}\label{eq:W1}
	P(W_2)=\sum_{j,k} \delta \left[W_2 - (E_k^{\tau}- E_j^0)\right] p_{jk}^\text{exp} p_j^0,
\end{equation}
where   $E_j^0$ and $E_k^{\tau}$ are the respective initial and final energy eigenvalues,  $p_j^0= \exp({-\beta_1 E_j^0})/Z^0$ is the initial thermal occupation probability with partition function $Z^0$ and $p_{j k}^\text{exp}= |\bra{j}U_\text{exp}(\tau)\ket{k}|^2$ denotes the transition probability between the instantaneous eigenstates $\ket{j}$ and $\ket{k}$ in time $\tau$ with  the corresponding unitary   $U_\text{exp}$.

Similarly, the probability density of the heat $Q=Q_3$ during the following heating step, given the expansion work $W_2$, is equal to the conditional distribution \cite{jar04},
  \begin{equation}\label{eq:Q2}
	P(Q|W_2)=\sum_{i,l} \delta \left[Q -(E_l^{\tau} - E_i^{\tau}) \right]p_{i l}^\text{hea} p_i^{\tau},
\end{equation}
where the occupation probability at time $\tau$ is $p_i^{\tau} = \delta_{k i}$ when the system is in eigenstate $\ket{k}$ after the second projective energy measurement.

The quantum work distribution for compression, given the expansion work $W_2$ and the heat $Q$, is additionally,
\begin{equation}\label{eq:W3}
	P(W_4|Q,W_2)= \sum_{r,m} \delta \left[ W_4 - (E_m^0 - E_r^{\tau}) \right] p_{r m}^\text{com} p_r^{\tau},
\end{equation}
with the  occupation probability $p_r^{\tau} = \delta_{r l}$ when the system is in eigenstate $\ket{l}$ after the third projective energy measurement. The transition probability $p_{r m}^\text{com}=|\bra{r}U_\text{com}(\tau)\ket{m}|^2$ is fully specified by the unitary time evolution operator for compression $U_\text{com}$.

The joint  probability of having certain values of $W_4$, $Q$ and $W_2$ during a cycle of the quantum engine now follows from the chain rule for conditional probabilities, $P(W_4,Q,W_2) = P(W_4|Q,W_2)P(Q|W_2)P(W_2)$ \cite{pap91}. Using Eqs.~\eqref{eq:W1}, \eqref{eq:Q2} and \eqref{eq:W3}, we find,
\begin{eqnarray}\label{eq:p_tot}
P(W_2,Q,W_4)& =& \sum_{j,k,l,m} \delta \left[W_2 - (E_k^{\tau}- E_j^0)\right] \nonumber \\
&\times& \delta \left[Q -(E_l^{\tau} - E_m^{\tau}) \right] \delta \left[ W_4 - (E_m^0 - E_l^{\tau}) \right]	\nonumber \\
&\times& p_{j}^{0}\,p_{jk}^{\text{exp}}\,p_{kl}^{\text{hea}}\,p_{lm}^{\text{com}} .
\end{eqnarray}
Introducing the total extracted work $W=-(W_2+W_4)$ work and integrating over all work values $W_2$ and $W_4$, the joint distribution for work and heat is given by,
\begin{equation}
	P(W,Q) = \int dW_2 dW_4 ~ \delta[W+(W_2+W_4)] P(W_2,Q,W_4).
\end{equation}
Using the explicit expression \eqref{eq:p_tot}, we finally obtain, 
\begin{eqnarray}\label{eq:p_tot_2}
P(W,Q) & = & \sum_{j,k,l,m}\Delta(W,j,k,l,m,\tau,\gamma)\Delta(Q,j,k,l,m,\tau,\gamma)\nonumber \\
 & &\times  p_{j}^{0}\,p_{jk}^{\text{exp}}\,p_{kl}^{\text{hea}}\,p_{lm}^{\text{com}}.
\end{eqnarray}
For ideal projective measurements, each spectral peak is infinitely sharp ($\gamma = 0$) and the two functions $\Delta$ associated with work and heat, $X=(W, Q)$,
are simply Dirac peaks, $\Delta(X,j,k,l,m,\tau,0)=\delta\left(X-x_{jklm} \right)$, with $w_{jklm} =E_{j}^{0} -E_{k}^{\tau}+E_{m}^{\tau} -E_{l}^{0}$  and $q_{jklm}=E_{l}^{\tau}-E_{m}^{\tau}$. 

In the experiment, each pair of energy measurements is effectively implemented using a Ramsey-like interferometric scheme \cite{dor13,maz13,bat14,bat15}. In this case,  spectral peaks have a finite width $\gamma$ and are well fitted by a Lorentzian distribution, $\Delta(X,j,k,l,m,\tau,\gamma)=1/\{\pi\gamma[1+ (X-x_{jklm})^2/\gamma^2]\}$ \cite{bat14,bat15}. This is the form we  consider in the present experiment.

\begin{figure*}
\includegraphics[width=.56\textwidth]{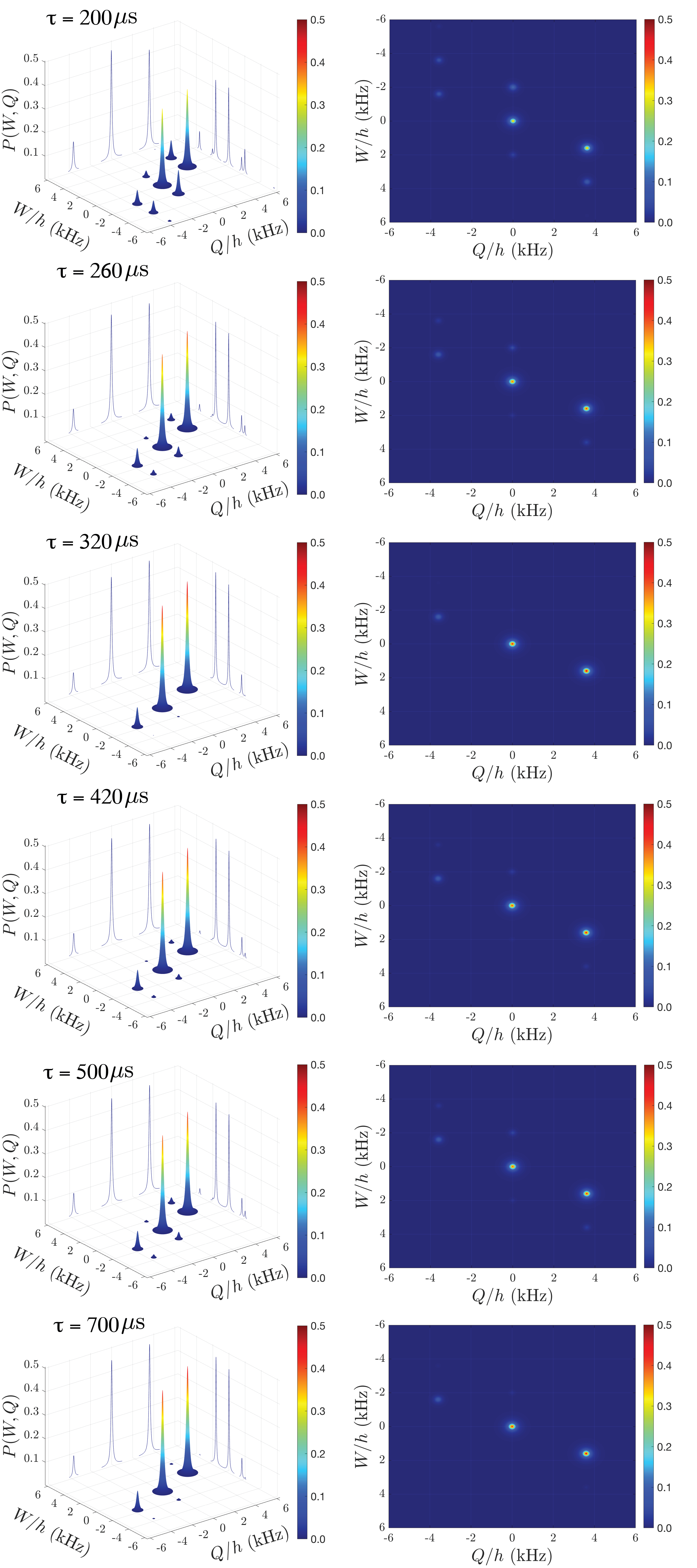}\caption{Reconstructed joint probability distribution for work and heat $P(W,Q)$, Eq.~\eqref{eq:p_tot_2},  for
the following values of the driving times, $\tau=200$, $260$, $320$,
$260$, $420$, $500$, and $700\,\mu$s (left), together with the corresponding density plots (right).}
\label{figs2}
\end{figure*}

\subsection{Joint distribution for work and heat - experiment}
We denote the instantaneous energy eigenstates of  the two-level system with energy gap $h\nu_i$ ($i= 1, 2)$  as $|\Psi_{\pm}^i\rangle$. The corresponding transition probabilities during expansion and compression strokes are accordingly given by 
\begin{equation}\label{eq:notrans}
	 |\bra{\Psi_{-}^{1}} U \ket{\Psi_{-}^{2}}|^2 =   |\bra{\Psi_{+}^{1}} U \ket{\Psi_{+}^{2}}|^2 = 1 - \xi,
\end{equation}
when there is no transition between states, and by 
\begin{equation}\label{eq:trans}
 |\bra{\Psi_{-}^{1}} U \ket{\Psi_{+}^{2}}|^2 =   |\bra{\Psi_{+}^{1}} U \ket{\Psi_{-}^{2}}|^2 =  \xi.
\end{equation}
when there is a change of state. The operator $U$ stands for the expansion or compression unitary. Adiabatic driving corresponds to  $\xi=0$.

Table \ref{table2} presents all the sixteen possible combinations for
energy transitions of the quantum Otto heat engine during one cycle, together with the
respective values of the  extracted random values of work and heat.

The reconstructed joint distributions $P(W,Q)$ are displayed in Fig.~\ref{figs2} for the following values of the driving time, $\tau=200$, $260$, $320$,
$260$, $420$, $500$, and $700\,\mu$s.

\subsection{Efficiency distribution}
The stochastic efficiency is defined as  $\eta= W/Q$. Its  distribution may be obtained from the joint distribution  $P(W,Q)$, Eq.~\eqref{eq:p_tot_2}, by integrating over $W$ and $Q$,
as 
\begin{eqnarray}
P(\eta)&=&\int dQdW~\delta\left(\eta-\frac{W}{Q}\right)P(W,Q) \nonumber\\
&=&\sum_{j,k,l,m} p_{j}^{0}\,p_{jk}^{\text{exp}}\,p_{kl}^{\text{hea}}\,p_{lm}^{\text{com}}L(w,q,\gamma,\eta)
\end{eqnarray}
with Lorentz-like peaks,
\begin{widetext}
\begin{eqnarray}
L(w,q,\gamma,\eta) & = & \frac{\gamma}{\pi^{2}\left(\gamma^{2}(\eta-1)^{2}+\eta^{2}q^{2}+2\eta qw+w^{2}\right)\left(\gamma^{2}(\eta+1)^{2}+\eta^{2}q^{2}+2\eta qw+w^{2}\right)}\nonumber \\
 & \times & \left\{ \gamma\left(-\gamma^{2}+\eta^{2}\left(\gamma^{2}+q^{2}\right)-w^{2}\right)\left(\log\left(\eta^{2}\right)+\log\left(\gamma^{2}+q^{2}\right)-\log\left(\gamma^{2}+w^{2}\right)\right)\right.\nonumber \\
 & + & \left.2\tan^{-1}\left(\frac{q}{\gamma}\right)\left(\eta^{2}q\left(\gamma^{2}+q^{2}\right)+2\eta w\left(\gamma^{2}+q^{2}\right)+q\left(\gamma^{2}+w^{2}\right)\right)\right.\nonumber \\
 & + & \left.2\tan^{-1}\left(\frac{w}{\gamma}\right)\left(\eta^{2}w\left(\gamma^{2}+q^{2}\right)+2\eta q\left(\gamma^{2}+w^{2}\right)+w\left(\gamma^{2}+w^{2}\right)\right)\right\} 
\end{eqnarray}
\end{widetext}
where we have dropped the indices  of $w$ and $q$ for better readability.

\begin{figure}[b!]
\centering \includegraphics[width=0.98\columnwidth]{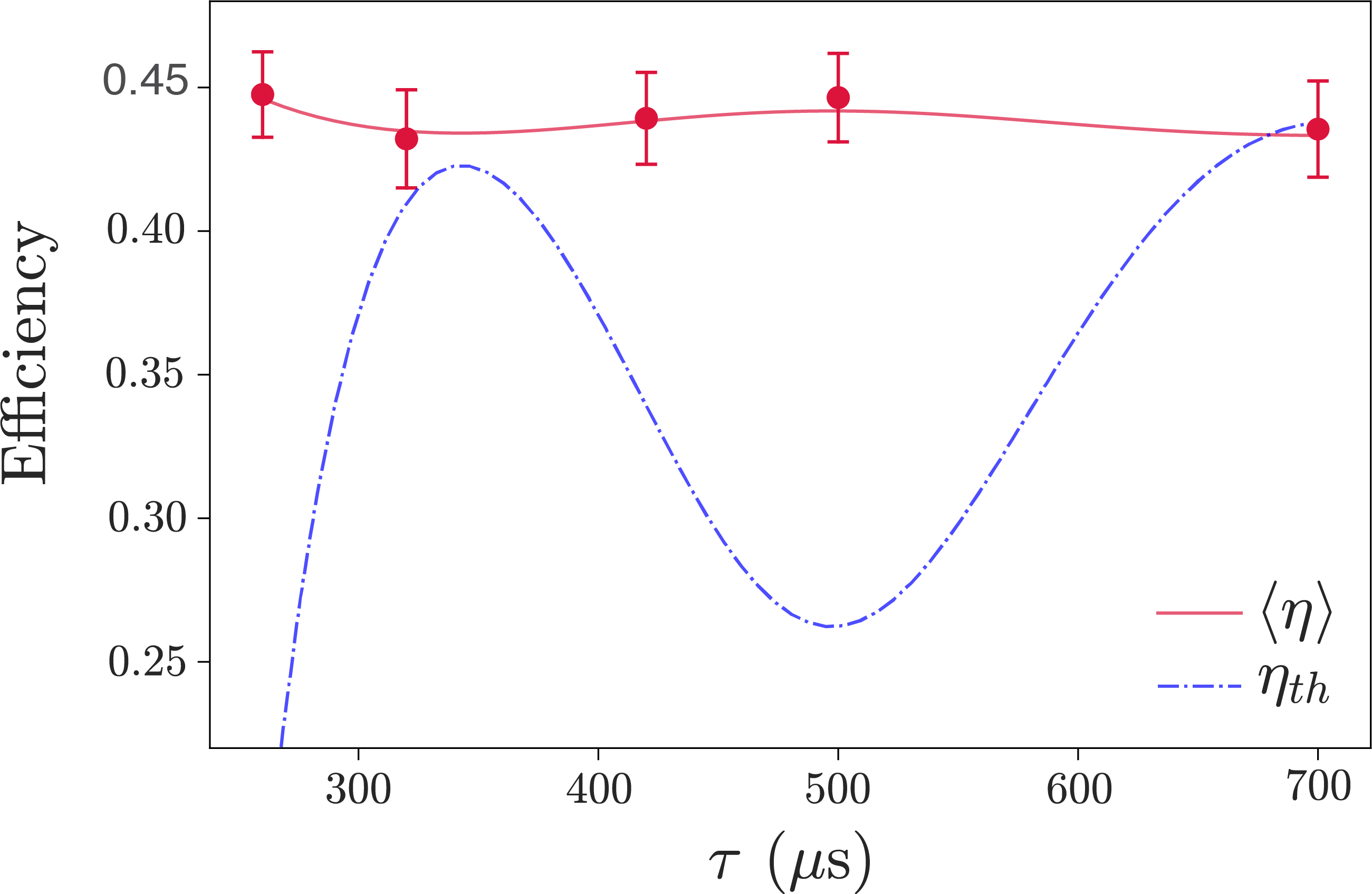} \caption{Comparison of the  microscopic mean efficiency $\langle \eta \rangle = \langle W/Q\rangle $ (experimental red dots) and the macroscopic efficiency $\eta_\text{th}= \langle W\rangle /\langle Q\rangle$ (simulated blue line) as a function of the driving $\tau$. The macroscopic efficiency increases as the adiabatic regime is approached, while the microscopic average efficiency decreases.}
\label{figs5}
\end{figure}
\subsection{Microscopic versus macroscopic efficiencies}
A comparison of the microscopic mean efficiency $\langle \eta \rangle = \langle W/Q\rangle $ and the macroscopic efficiency $\eta_\text{th}= \langle W\rangle /\langle Q\rangle$ is displayed in Fig.~\ref{figs5} as a function of the driving time $\tau$. The macroscopic efficiency $\eta_\text{th}$ (simulated blue line) increases as the adiabatic regime is approached and irreversible losses induced by quantum friction are reduced. By contrast, the microscopic mean efficiency $\langle \eta \rangle$ (experimental red dots) decreases near the adiabatic regime. It is hence larger for nonadiabatic driving. This counterintuitive behavior is due to the presence of  peaks above $\eta_\text{th}$ and, thus, to random events that violate the macroscopic second law law of thermodynamics.

\subsection{Detailed fluctuation relation}

\begin{figure*}
\includegraphics[width=.85 \textwidth]{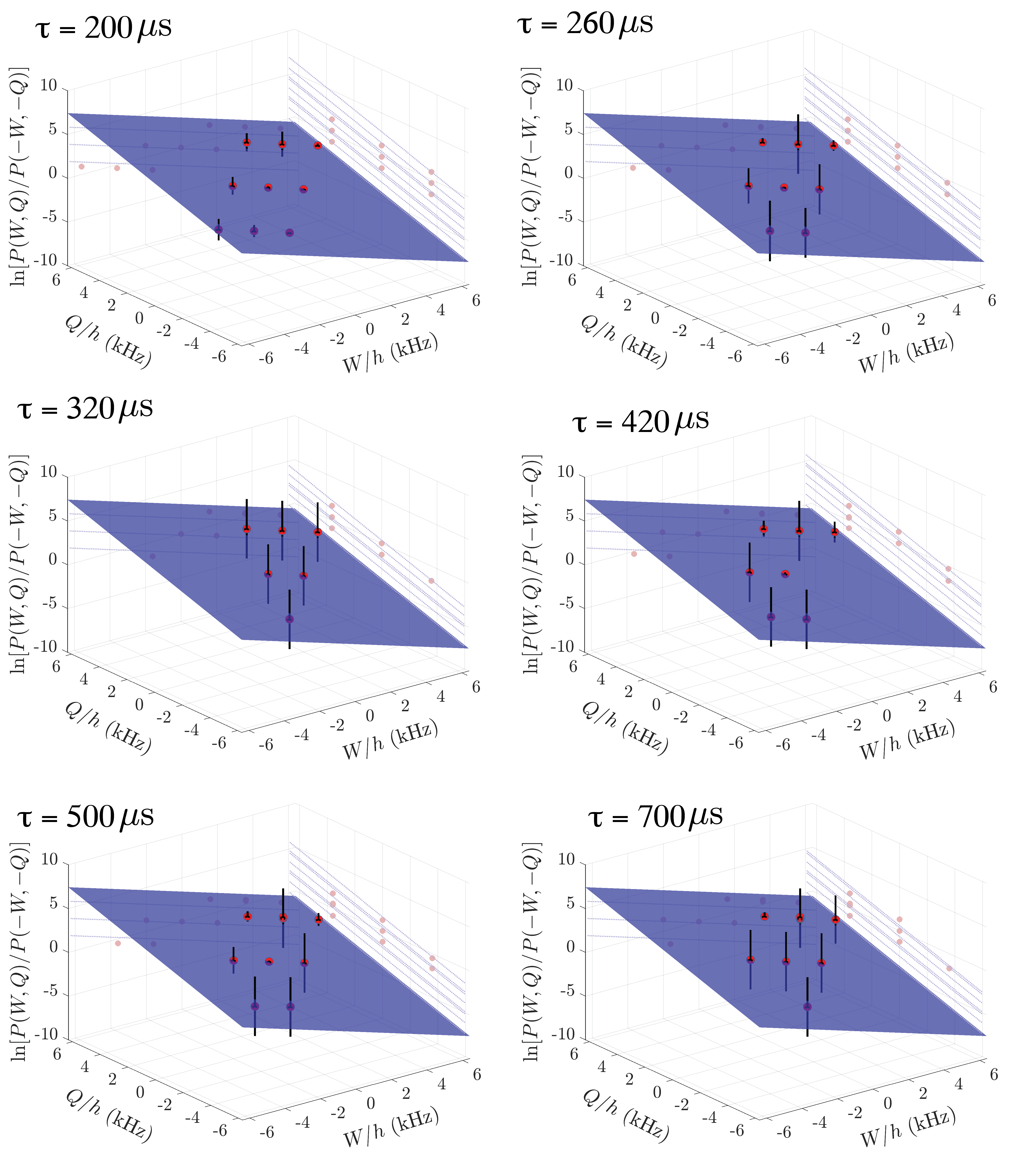}
\caption{Experimental verification of the detailed quantum fluctuation relation for work and heat for the
following values of the driving time, $\tau=200$, $260$, $320$, $260$,
$420$, $500$, and $700\,\mu$s.}
\label{Sup03}
\end{figure*}

A test of the detailed quantum fluctuation relation, 
\begin{equation}
\frac{P(W,Q)}{P(-W,-Q)}=e^{-\Delta\beta Q-\beta_{1}W}.
\end{equation}
was presented in the main text for the driving time $\tau=200\,\mu$s. Figure \ref{Sup03} exhibits similar tests for $\tau=200$,
$260$,  $260$, $320$, $420$, $500$, and $700\,\mu$s, showing that the fluctuation theorem for work and heat is obeyed for all the driving times realized in the experiment.

\subsection{Numerical simulations} 

The numerical simulation of the experiment was implemented using a python-based code (in-house developed) and QuTiP
(the Quantum Toolbox in Python) package \cite{Python}. We effectively
simulated the finite-time quantum Otto cycle described in the main
text with the thermalization strokes being solved using the theoretical
thermalization of a qubit with a Markovian thermal reservoir in terms
of the Bloch vector components \cite{Chakraborty2019}. The time-dependent unitary dynamics of the energy gap expansion and compression strokes were solved numerically. In order to obtain the theoretical transition probability, we  ran the simulation from $\tau=100\mu s$ to $\tau=700\mu s$ considering
$50$ time steps, which was sufficient to generate smooth curves for
the theoretical quantities and for the confirmation of the  quantum fluctuation relations.\\

\textit{Acknowledgements.} We acknowledge financial support from the Federal
University of ABC (UFABC), the Brazilian National Council for Scientific
and Technological Development (CNPq), the Brazilian Federal Agency for
Support and Evaluation of Graduate Education (CAPES),  the S\~ao
Paulo Research Foundation (FAPESP) (Grant number 19/04184-5) and the German Science Foundation (DFG) (Project FOR 2724). This research was performed as
part of the Brazilian National Institute of Science and Technology
for Quantum Information (INCT-IQ). We also thank the Multiuser Central
Facilities of UFABC.

\end{document}